\documentclass[twocolumn,nolinenumbers]{aastex631}
\usepackage{graphicx}
\usepackage{dcolumn}
\usepackage{bm}
\usepackage{amssymb}
\usepackage{amsmath}

\begin{document}

\title{High-energy Neutrino Emission from NGC 1068 by Outflow-cloud Interactions}

\author{Yong-Han Huang}
\affiliation{Department of Astronomy, School of Physics, Huazhong University of Science and Technology, Wuhan 430074, China}

\author[0000-0003-4976-4098]{Kai Wang}
\affiliation{Department of Astronomy, School of Physics, Huazhong University of Science and Technology, Wuhan 430074, China}

\author[0009-0007-0717-3667]{Zhi-Peng Ma}
\affiliation{Department of Astronomy, School of Physics, Huazhong University of Science and Technology, Wuhan 430074, China}

\correspondingauthor{Kai Wang}
\email{kaiwang@hust.edu.cn}

\begin{abstract}
As the hottest high-energy neutrino spot, NGC 1068 has received much attention in recent years. Here we focus on the central region of the active galactic nuclei (AGN) and propose an outflow-cloud interaction model that could probably explain the observed neutrino data. Considering the accretion process adjacent to the central supermassive black hole (SMBH) of NGC 1068, strong outflows will be generated, which will likely interact with surrounding clouds floating in the corona region. Particles carried by the outflow will be accelerated to very high energy by the shocks forming during the outflow-cloud interactions. For the accelerated high-energy protons, $p\gamma$ interactions with the background photon field of the corona and disk and $pp$ interaction with the surrounding gas will produce considerable high-energy $\gamma$-rays and neutrino. However, because of the extremely dense photon fields in the corona and disk, the newly generated $\gamma$-rays will be significantly attenuated through the $\gamma\gamma$ absorptions. In our scenario, the expected GeV-TeV $\gamma$-ray emission will be suppressed to a much lower level than the neutrino emission, consistent with the observational characteristics of NGC 1068, while the generated 1-30\,TeV neutrino flux can fit the IceCube data very well.    

\end{abstract}


\section{Introduction} \label{sec:intro}
Active galactic nuclei (AGNs) are known as powerful high-energy astrophysical sources. Ultra-fast outflow can be continually generated from the central region of the AGN with a mildly relativistic velocity of about 0.03-0.3 c~\citep{peretti2023gamma}. 
NGC 1068 is a typical Seyfert II galaxy with an AGN~\citep{1997Ap&SS.248....9B} and is found to be the hottest neutrino spot in the 10-year survey data of IceCube with 4.2$\sigma$ confidence level~\citep{icecube2022evidence}. NGC 1068 is about 14.4 Mpc away from us with a supermassive black hole (SMBH) in the center covered by thick gas and dust~\citep{garcia2016alma, gamez2022thermal}. According to the hard X-ray detection of NGC 1068, ~\cite{matt1997hard} suggests that it is Compton thick and difficult to detect the existence of the outflow. However, as indicated in~\cite{mizumoto2019kinetic}, it is still possible that NGC 1068 has ultra-fast outflow. This energetic outflow can accelerate particles to PeV or even EeV~\citep{peretti2023diffusive} and interact with the AGN photon field through hadronic processes, producing high-energy neutrinos and $\gamma$-rays.
For the hadronic processes, the production of very-high-energy (VHE) neutrinos always coincides with the emission of $\gamma$-rays, and the luminosity of the $\gamma$-ray emissions is comparable to that of the neutrinos~\citep{gamez2022thermal}. However, it is found that its TeV neutrino flux for NGC 1068 is at least an order of magnitude larger than that of $\gamma$-rays~\citep{icecube2022evidence}. Thus, it is believed that the accompanying GeV-TeV $\gamma$-rays with neutrinos are highly obscured.

The neutrino production region is considered at the central corona region in the vicinity of SMBH to absorb the accompanying GeV-TeV $\gamma$-rays (e.g., in \cite{2020ApJ...891L..33I,inoue2022high,2021Galax...9...36I,2021ApJ...922...45K,Anchordoqui:2021C3,2022ApJ...939...43E,PhysRevLett.125.011101,2022ApJ...941L..17M,2024ApJ...961L..34M}) since X-ray emissions in the corona region have been suggested to be very bright~\citep{Bauer_2015,10.1093/mnrasl/slv178}. Besides, the high-energy neutrinos produced by outflow-torus~\citep{inoue2022high} or jet-ISM (interstellar medium)~\citep{fang2023high} interactions have been explored as well.

\begin{figure}
	\centering
	\includegraphics[width = 0.99\linewidth , trim = 0 0 0 0,clip]{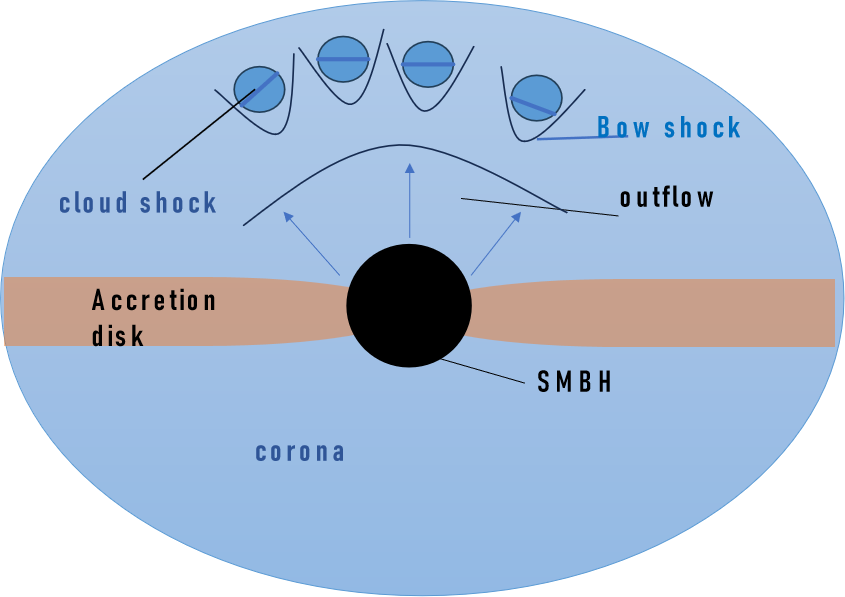}
	\caption{The schematic outflow-cloud interaction model. The outflow launched from the SMBH-disk system collides with the cloud in the corona region. Particles undergo acceleration by the bow shock and cloud shock during the outflow-cloud interaction. Then high-energy neutrinos can be generated by hadronic processes with the surrounding photon fields or gas.}
\label{fig 1}	
\end{figure}

Here we propose the outflow-cloud interaction model (shown in Fig.~\ref{fig 1}) to explain the neutrino emission observed by Icecube. This kind of outflow with ultra-fast velocity has been discovered in many AGNs and quasars by analyzing their X-ray spectra~\citep{chartas2002chandra,chartas2003xmm,chartas2009confirmation, pounds2003fe, dadina2005x,markowitz2006fe, braito2007relativistic, cappi2009x,  reeves2009compton, giustini2011variable, gofford2011broad, lobban2011contemporaneous, dauser2012spectral}. Outflow launched by the SMBH-disk system passes through the corona region, where it is likely to interact with dense dark clouds around the central SMBH in the corona region. When outflow collides with the cloud, a bow shock can be produced outside the cloud, and it will significantly accelerate the particles to rather high energy through the diffuse shock acceleration (DSA) mechanism. Considering the high-density photon field existing in the central region of the AGN, those efficiently accelerated particles will be able to have $p\gamma$ interactions with X-ray and ultraviolet photon fields produced by the corona and disk respectively. Also, those accelerated particles may have $pp$ interactions with the outflow matters to produce $\gamma$-rays and neutrinos at the same time.
 
 
\section{Dynamics and physical parameters for outflow-cloud interaction} 
After outflows from SMBH collide with the surrounding clouds, there will be two types of shock~\citep{mckee1975interaction}, namely, a bow shock (BS) and a cloud shock (CS). Both shocks will accelerate particles, nevertheless, BS only exists outside the cloud while CS can sweep through the whole cloud. Here we provide two sets of parameters for the $pp$ dominant case and the $p\gamma$ dominant case. First, in Case 1 (Case 2) we assume that the outflows are spherically symmetric with a velocity $V_{0}$ of 0.3 c (0.2 c) and a kinetic luminosity $L_{\rm kin}$ = $5\times10^{44}\,\rm erg\, s^{-1} $ ($10^{46}\,\rm erg\, s^{-1} $). According to \cite{inoue2022high}, the characteristic radius of the corona is $10^{14} $ cm, so we assume that the clouds are situated around $5\times10^{13}\,\rm cm$ away from SMBH.  Then the kinetic luminosity can be written as $L_{\rm kin}=2\pi r_{0} ^{2} \rho _{0}V_{0} ^{3}  $, where $r_{0}$ is the distance from the SMBH and $\rho _{0}$ is the corresponding mass density of the outflow at the distance of $r_{0}$. So, it is easy to derive the number density of outflow in the typical cloud region, i.e., $n_{0}=\frac{\rho _{0} }{m_{H}  } \sim 3\times 10^{10}$$\left ( \frac{L_{\rm kin} }{5\times10^{44}\,\mathrm{erg \, s^{-1} } }  \right ) \left ( \frac{V_{0} }{0.3c}  \right )^{-3\ } \left ( \frac{r_{0} }{5\times10^{13}\,\mathrm{cm} } \right )^{-2}\,\rm cm^{-3}$ ($2\times10^{12} \,\rm cm^{-3}$), with $m_{H}$ the mass of hydrogen atom. All adopted parameters have been listed in Table~\ref{table:joint}.

According to \citep{mckee1975interaction,mou2021years}, we have a relationship between the outflow velocity $V_{0} $ and the cloud shock velocity of $V_{c} =\chi ^{-0.5} V_{0} $, where $\chi =\frac{n_{c} }{n_{0} }$ and $n_{c}$ is the number density of cloud particles, which should be higher than the outflow number density, otherwise, the cloud will be destructed by outflow soon after they crash and there will be little time to accelerate particles. Given this, $n_{c}$ is set to $10^{12} \,\rm cm^{-3}  $ ($10^{14} \,\rm cm^{-3} $), thus $\chi \simeq 40$ (60).

The lifetime of the cloud can directly affect the efficiency of DSA process as the particle acceleration will cease when the cloud is destructed by the outflow and shocks. As discussed in \cite{klein1994hydrodynamic}, the lifetime of the cloud ($T _{\rm cloud}$) is comparable to the timescale of CS that sweeps the cloud, that is, $T_{\rm cloud} =\frac{r_{c} }{V_{c} } =\frac{r_{c} }{V_{0} } \chi ^{0.5} \sim 30\,\rm s$ (60s) with ${r_{c}=5\times10^{10} }$cm the radius of cloud. Next, according to \cite{drury1983introduction}, a particle (in most cases a proton) with charge number Z and energy $ E_{p} $ can be accelerated by BS on a timescale of $T_{\rm BS} \approx \frac{8}{3} \frac{cE_{p} }{ZeBV_{0}^{2}  } $ and by CS in a timescale of $T_{\rm CS} \approx \frac{8}{3} \frac{cE_{p} }{ZeBV_{c}^{2}  } $, where $e$ is the electron charge and $B=10\,\rm G$ (25G) is the magnetic field strength. For simplicity, we assume that the outflow and clouds share the same magnetic field. In addition, we have considered the possibly existing $pp$ interactions. We use an approximate equation to estimate the $pp$ interaction timescales: in the BS region, it is $t_{pp,\rm BS} \simeq \frac{1}{cn_{0} \sigma _{pp} } $, and in the CS region (inside the cloud), it is $t_{pp,\mathrm{CS}} \simeq \frac{1}{cn_{c} \sigma _{pp} } $, where $\sigma _{pp}\approx 30\,\rm mb$ is the $pp$ cross-section. 

The emissions from the corona and disk are the two main components taken into account in our calculation for the $p\gamma$ interaction. We adopt the disk photon spectral energy distribution (SED) as a diluted blackbody spectrum with a typical photon energy of 32 eV, and total luminosity $L_{\rm disk}=10^{45}\,\rm erg \, s^{-1} $\citep{woo2002active,zaino2020probing}. Given the observation of NuSTAR and XMM-Newton~\citep{bauer2015nustar,marinucci2015nustar}, we choose $L_{\rm X} =1.4\times 10^{44} \rm erg\, s^{-1} $ ($2-10$ keV band) as the luminosity of corona X-ray photon field and a index of -2 with exponential cutoff energy of 128 keV for SED.

\begin{figure}
	\centering
	\includegraphics[width = 1\linewidth , trim = 0 0 0 0,clip]{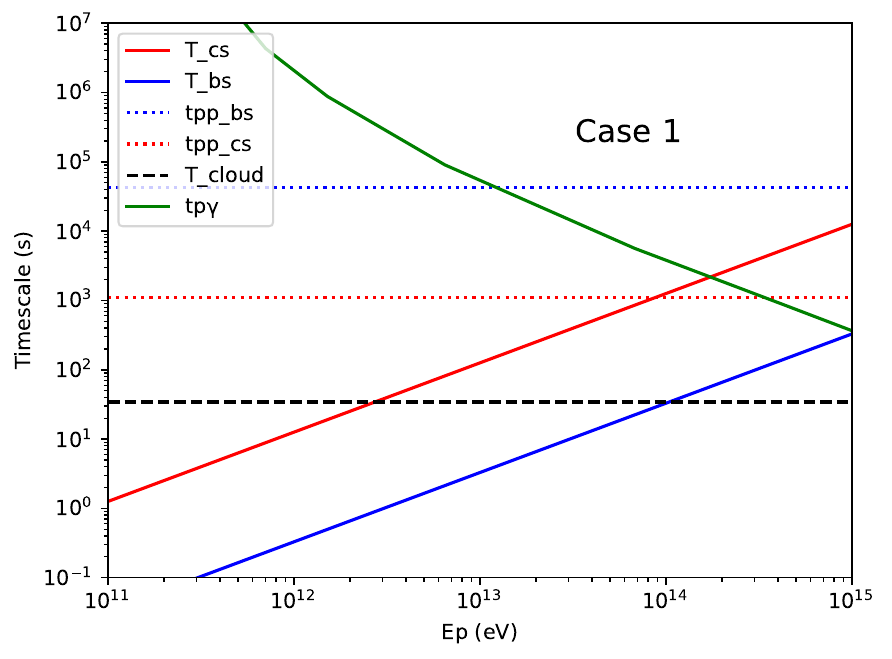}
        \includegraphics[width = 1\linewidth , trim = 0 0 0 0,clip]{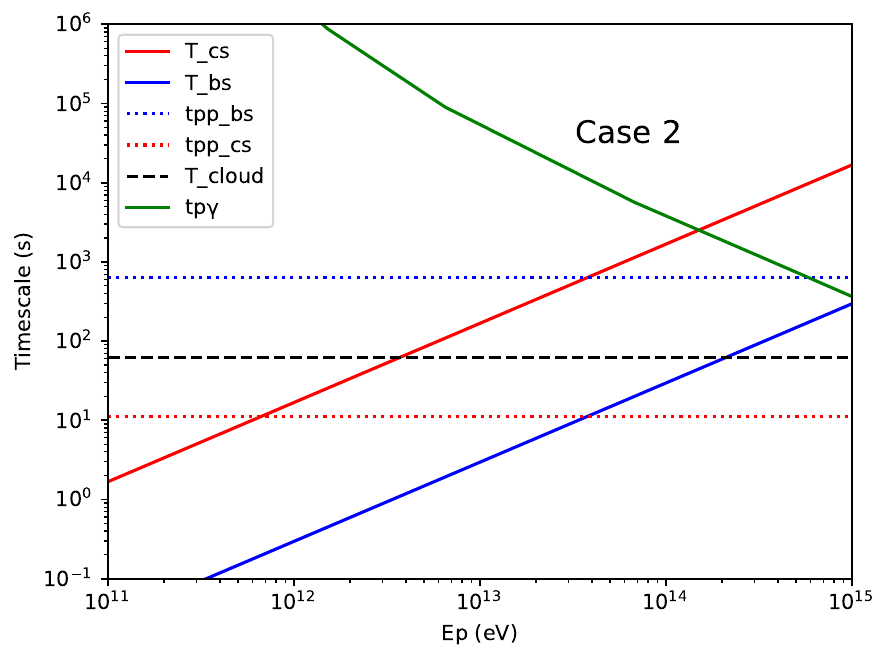}
	\caption{\textit{Upper panel:} Case 1; \textit{Bottom panel:} Case 2. The blue, red, and green solid lines represent the acceleration timescales of the BS and CS, and the $p\gamma$ cooling timescale respectively. The dotted lines show the $pp$ cooling timescales including at the BS site and inside the cloud, and the black dashed line is the lifetime of the cloud. All parameters are listed in Table~\ref{table:joint}.}
\label{fig 2}	
\end{figure}

Therefore, the $p\gamma$ timescale can be calculated by~\citep{murase2007high}
\begin{equation}
\	t_{p\gamma }^{-1}=\frac{c}{2\gamma _{p}^{2} } \int_{\bar{\varepsilon } _{th} }^{\infty}
\sigma _{p\gamma } \left ( \bar{\varepsilon }  \right )\kappa_{p\gamma } \left ( \bar{\varepsilon }  \right )
\bar{\varepsilon }d\bar{\varepsilon }\int_{\frac{\bar{\varepsilon }}{2\gamma _{p} } }^{\infty}\varepsilon ^{-2 } \frac{dn}{d\varepsilon }d\varepsilon,
\end{equation}
where $\sigma _{p\gamma } \left ( \bar{\varepsilon }  \right )$ and $\kappa_{p\gamma } \left ( \bar{\varepsilon }  \right )$ represent the cross section and the inelastic parameter respectively. We set $\sigma _{p\gamma } \left ( \bar{\varepsilon }  \right )$ to 0.5 mb for simplicity and take $\kappa_{p\gamma } \left ( \bar{\varepsilon }  \right )$ as described in \cite{stecker1968effect}. $\gamma _{p} $ is the Lorentz factor of the proton, $\bar{\varepsilon }$ is the photon energy in the proton rest frame, and the threshold energy $\bar{\varepsilon } _{th}\simeq  145\,\rm MeV$. Now we can plot all the timescales in Fig.~\ref{fig 2}. Interestingly, in Case 1, we find that the BS $pp$ interaction efficiency surpasses that of $p\gamma$ interaction at the energy below $\sim$10 TeV, and the $p\gamma$ interaction dominates in energy above $\sim$10 TeV. In Case 2, due to the high proton luminosity, the $pp$ interaction efficiency is significantly higher than the $p\gamma$ interaction and dominates in the energy of $\lesssim7\times 10^{14}\,$eV. Now one can get the accelerated maximum energy by the BS using $T_{\rm BS}=\min(T_{\rm cloud},t_{pp,\rm BS},t_{p\gamma})$, says, $E_{p,\rm max,bs}\simeq 100\left ( \frac{B}{10\,\rm G}  \right ) \left ( \frac{V_{0} }{0.3c}   \right )^{2}  \left ( \frac{T_{\rm cloud} }{30\,\rm s}  \right ) \,\rm TeV $ $(200\,\rm TeV)$ and by the CS using $T_{\rm CS}=\min(T_{\rm cloud},t_{pp,\rm CS},t_{p\gamma})$, says, $E_{p,\rm max,cs}\simeq 3\left ( \frac{B}{10\,\rm G}  \right ) \left ( \frac{V_{c} }{1\times10^9 \,\rm cm\,s^-1}   \right )^{2}  \left ( \frac{t_{pp,\rm CS} }{10\,\rm s}  \right ) \,\rm TeV $ $(0.6\,\rm TeV)$. As seen in Fig.~\ref{fig 2}, the maximum proton energy is typically determined by the lifetime of the cloud. 

As the maximum proton energy $E_{p,\rm max,cs}$ accelerated by CS is quite small and outside of the interesting energy region, we neglect the accelerated protons by CS and only focus on the BS acceleration. Another reason to neglect the CS is that the CS energy converted from the kinetic energy of the outflow is small since the energy ratio between the CS and BS is proportional to $\chi^{-0.5} \simeq 0.16$ ($\simeq 0.13$ for Case 2)~\citep{mou2021years}. In addition, different with \cite{wu2022could}, we assume the most accelerated protons by BS are advected away with the downstream shocked materials of BS without effectively entering the cloud. Therefore, the $pp$ interactions inside the cloud are neglected and only the $pp$ interactions between the accelerated protons by BS and the outflow materials are taken into account.

\section{Hadronic process } 
Next, we calculate the high-energy neutrino and $\gamma$-ray productions by the $pp$ and $p\gamma$ interactions. 
The accelerated proton distribution is adopted as a power law with the maximum energy exponential cutoff, i.e.,
\begin{equation}
\	\frac{dN\left ( E_{p}  \right ) }{dE_{p} } =AE_{p} ^{\Gamma _{p}}e^{\left ( -\frac{E_{p} }{E_{p,\max}}  \right ) } ,
\end{equation}
where the index $\Gamma _{p}=-2 $  is adopted. Since the $pp$ and $p\gamma$ process happens around the site of outflow-cloud interaction, the normalization factor $A$ can be calculated by 
\begin{equation}
\	L_{p}=\alpha \beta L_{kin} =2\pi R_{\rm 0}^{2} V_{0}\int E_{p} \frac{dN\left ( E_{p}  \right ) }{dE_{p} }dE_{p}   
\end{equation}
with the effective proton luminosity $L_{p}=5\times10^{43} \,\rm erg\,  s^{-1}$, the covering factor (fraction of coverage) of the cloud  $\alpha\sim0.3$  and the energy transferring rate (energy fraction that can be used to accelerate particles by BS) $\beta\sim0.3$.

By using the method in \cite{kelner2008energy}, we get the spectrum of the $\gamma$-ray and neutrinos produced by the $p\gamma$ interaction.
\begin{equation}
	\frac{dN_{l } }{dE_{l } }= \int f_{p}(E_{p} ) f_{ph} (\epsilon )\Phi _{l}\left ( \eta ,x \right )\frac{dE_{p}}{E_{p} } d\epsilon    
\end{equation}
where subscript $l=\nu$ or $\gamma$, $\Phi _{l}$ is a specific function form (see \cite{kelner2008energy} for details), $\eta $ is defined by $\eta=\frac{4\epsilon E_{p} }{m_{p}^{2}c^{4}   } $, $x=\frac{E_{l} }{E_{p} } $, and $f_{p},f_{ph}$ are distribution functions of proton and target photon respectively. To calculate the neutrino and $\gamma$-ray spectrum of $pp$ interaction, we refer to the formula (62), (66) and (58) in ~\cite{kelner2006energy}.

\begin{figure}
	\centering
	\includegraphics[width = 1\linewidth , trim = 0 0 0 0,clip]{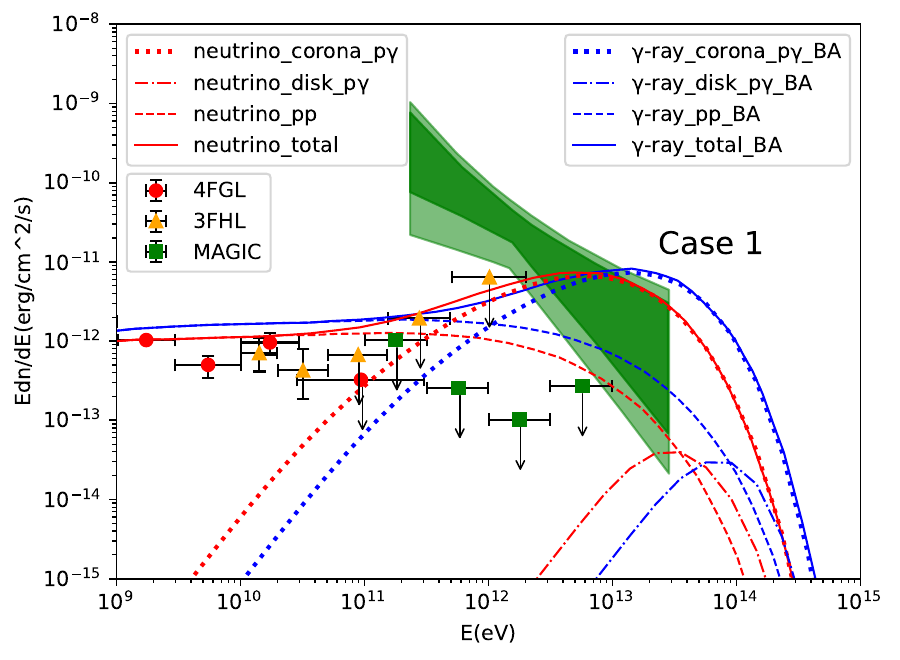}
        \includegraphics[width = 1\linewidth , trim = 0 0 0 0,clip]{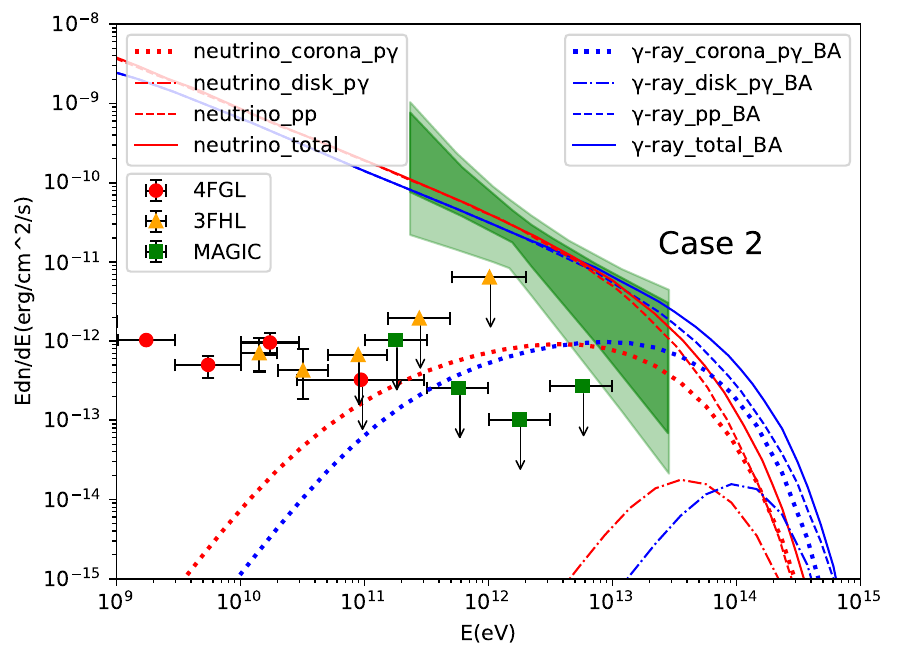}
	\caption{(“BA” refers to “before attenuation”). \textit{Upper panel:} The red and blue dotted lines are neutrino and $\gamma$-ray from the $p\gamma$ interaction with the corona radiation, the red and blue dash-dotted lines are neutrino and $\gamma$-ray from the $p\gamma$ interaction with the disk radiation, the red and blue dashed lines are neutrino and $\gamma$-ray from the $pp$ interaction with the outflow gas. The blue and red solid lines are total $\gamma$-ray and neutrino flux from both $pp$ and $p\gamma$ interactions. For comparison, we also put the $\gamma $-ray observation data from 4FGL, 3FHL, and MAGIC in the figure. \textit{Bottom panel:} This panel shows the case in which the pp interaction dominates. Note that the theoretical $\gamma$-rays after absorptions in our scenario are invisible in the figure, and the $\gamma$-rays before attenuations are plotted for reference (see more details in Sect.~\ref{attenuation}). The dark green and light green shaded regions represent the observed muon and anti-muon neutrino flux with a confidence level of 3$\sigma $ and 2$\sigma $ respectively~\citep{aartsen2020time}. All adopted parameters are listed in Table~\ref{table:joint}.}
\label{fig 3}	
\end{figure}

After putting different parameters of corona and disk into the above equations, we obtain the total flux of $\gamma$-rays and neutrinos for different cases (see Fig.~\ref{fig 3}).
From the upper panel of Fig.~\ref{fig 3}, we can find out that the $p\gamma$ process in the corona contributes most of the observed neutrino flux and fits well with the IceCube data ranging from 1-30 TeV. At lower energies of neutrinos (below 1 TeV), the $pp$ interaction between the accelerated protons and the outflow materials can contribute more than that of the $p\gamma$ process. As shown in Case 2 (the bottom panel of Fig.~\ref{fig 4}), where the $pp$ interaction always dominates in the corona, the observed neutrino flux data could be nicely fit by the single $pp$ process. Also, the corona plays the main role in generating high-energy $\gamma$-rays, while the disk has little influence on that. Interestingly, the corona itself is opaque to the high-energy $\gamma$-ray produced inside it because of the large optical depth as presented in the next Section. In our scenario, the GeV-TeV $\gamma$-rays are significantly suppressed due to the quite large optical depths induced by the corona and disk emissions. Thus, the explanation of GeV-TeV $\gamma$-rays has to be invoked by other physical mechanisms, e.g., from the starburst region~\citep{2022ApJ...939...43E}.

\begin{table*}[ht]
    \centering
    \caption{The adopted parameters for Case 1 and Case 2 fittings}
    \label{table:joint}
    \begin{tabular}{c|c|c|c} 
    \hline
        \textbf{Names} & \textbf{Symbols} & \textbf{Case 1} & \textbf{Case 2} \\
        \hline  
        Outflow velocity & $v_{0}$ & $0.3c$ & $0.2c$ \\
        \hline  
        Cloud distance & $r_{0}$ & $5\times 10^{13}\,\rm cm$ & $5\times 10^{13}\,\rm cm$ \\
        \hline      
        Outflow density & $n_{0}$ & $3\times10^{10}\,\rm cm^{-3}$ & $2\times10^{12}\,\rm cm^{-3}$ \\
        \hline  
        Cloud density & $n_{c}$ & $10^{12}\,\rm cm^{-3}$ & $10^{14}\,\rm cm^{-3}$ \\
        \hline  
        Cloud radius & $r_{c}$ & $5\times 10^{10}\,\rm cm$ & $5\times 10^{10}\,\rm cm$ \\
        \hline  
        Magnetic strength & $B$ & $10\,\rm G$ & $25\,\rm G$ \\
        \hline 
        Proton luminosity & $L_{\rm p}$ & $5\times10^{43}\,\rm erg \, s^{-1}$ & $10^{45}\,\rm erg \, s^{-1}$ \\
        \hline 
        Proton maximum energy & $E_{p,\max}$ & $10^{14}\,\rm eV$ & $2\times 10^{14}\,\rm eV$ \\
        \hline

        Proton spectral index & $\Gamma _{\rm p} $ & -2 & -2.7 \\
        \hline      
        Photon spectral index & $\Gamma _{\rm ph} $ & -2 & -2 \\
        \hline
        Kinetic luminosity & $L_{\rm kin}$ & $5\times10^{44} \,\rm erg \, s^{-1}$ & $10^{46} \,\rm erg \, s^{-1}$ \\
        \hline      
        Corona luminosity & $L_{\rm cor}$ & $1.4\times10^{44}\,\rm erg \, s^{-1} $ & $1.4\times10^{44}\,\rm erg \, s^{-1} $ \\
        \hline      
        Disk luminosity & $L_{\rm dis}$ & $10^{45}\,\rm erg \, s^{-1}$ & $10^{45}\,\rm erg \, s^{-1}$ \\
        \hline      
        BLR luminosity & $L_{\rm BLR}$ & $10^{44}\,\rm erg \, s^{-1}$ & $10^{44}\,\rm erg \, s^{-1}$ \\
        \hline      
        BLR radius & $R_{\rm BLR}$ & $5\times10^{16}\,\rm cm$ & $5\times10^{16}\,\rm cm$ \\
        \hline      
        Covering factor & $\alpha$ & 0.3 & 0.3 \\
        \hline      
        Energy transfer efficiency & $\beta$ & 0.3 & 0.3 \\  
        \hline
    \end{tabular}
\end{table*}

\section{Attenuation of gamma rays} \label{attenuation}
\begin{figure}
	\centering
	\includegraphics[width = 1\linewidth , trim = 0 0 0 0,clip]{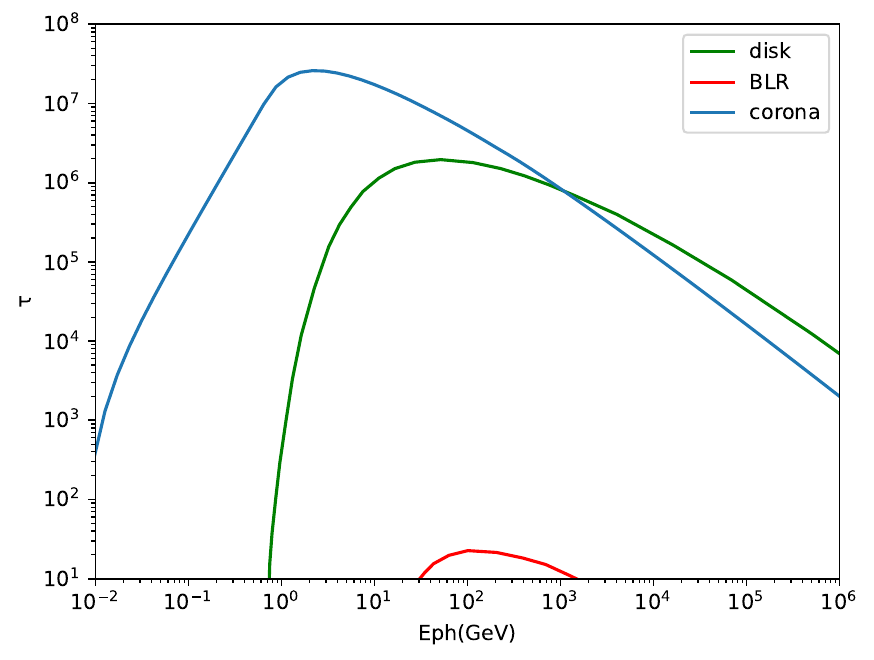}
	\caption{The blue, green, and red lines show the optical depths of the corona, disk, and BLR, respectively, for Case 1.  }
\label{fig 4}	
\end{figure}

Since the absorption by the background photon fields of the corona, disk, and broad line region (BLR) will affect the $\gamma$-ray produced in the central region, we also calculate the optical depths of different regions as shown in Fig.~\ref{fig 4}. 

In our calculations, the cross-section of $\gamma\gamma$ attenuation is adopted as~\citep{mou2021years}
\begin{equation}
	\sigma _{\gamma \gamma }\left ( E_{\gamma  }, \nu , \alpha  \right )  =
 \frac{3\sigma _{T} }{16}
\left ( 1-\beta ^{2}  \right )\left [ 2\beta \left ( \beta ^{2} -2 \right )+\left ( 3-\beta ^{4}  \right ) \ln\frac{1+\beta }{1-\beta }   \right ],   
\end{equation}
where $\beta =\sqrt{1-\frac{2m_{e}^{2}c^{4}   }{h\nu E_{\gamma } \left ( 1-\cos \alpha  \right ) } }$ and $\alpha$ is the colliding angle of two photons ($E_{\gamma  }$ and $h\nu$) in the lab frame. $\sigma _{T} $ is the Thomson scattering section. Notice that we treat the background photon field of corona, disk, and BLR as anisotropic, and assume that it is spherically symmetric for the SMBH. Thus, it is convenient to establish a spherical coordinate and make $\cos \alpha=\frac{\left ( r^{2}+R^{2}-r_{t }^{2}     \right ) }{2rR}$, where $R= \sqrt{r^{2} +r_{t} ^{2}+2rr_{t} \cos \theta}  $ with $r_{t}$ the distance from the SMBH to where the $\gamma$-ray is generated, $\theta$ the angle between the momentum of $\gamma$-ray and the vector pointing from SMBH to the place where the $\gamma$-ray is generated and $r$ the distance a $\gamma$-ray photon travels until it annihilate with other photons. The high-energy $\gamma$-rays produced by the $p\gamma$ or $pp$ interactions are supposed at $r_{t}=5\times 10^{13}\,\rm cm$, namely, the position of outflow-cloud interaction, since the typical timescales of hadronic interactions shown in Fig.~\ref{fig 2} are faster than $r_{t}/c$. For the low-energy photon fields, the disk and corona emissions are treated as filling the whole space and scaled by the actual radius $R$ and their luminosities. We also consider the low-energy BLR photon field. For the BLR radiation field, we assume its luminosity to be  $L_{\rm BLR} = 10^{44} \,\rm erg\, s^{-1} $~\citep{muller2020radiation} and the typical radius $R_{\rm BLR}$ to be $5\times10^{16}\,\rm cm$, with a diluted black body spectrum and a typical energy of 10 eV~\citep{abolmasov2017gamma}. The BLR emission is assumed to attenuate the high-energy $\gamma$-rays from its typical radius, so one has $r_t=5\times 10^{16}\,\rm cm$.

Finally, we can derive the optical depth in the form of 
\begin{equation}
	 \tau _{\gamma \gamma } =\iiint \sigma _{\gamma \gamma } \left ( E_{\gamma} ,\nu ,\alpha   \right ) n_{ph}\left ( \nu ,r,\theta  \right )\frac{\sin \theta }{2} d\nu drd\theta.
\end{equation}
By putting the corresponding background photon field ($n_{ph}$) of the different regions (BLR, corona, disk) into the above equation, one can get the optical depths, and the results are plotted in Fig.~\ref{fig 4}. 
One can see from Fig.~\ref{fig 4} that corona and disk photon fields dominate in the absorption of produced $\gamma$-rays, while the BLR can be ignored in comparison. The $\gamma\gamma$ absorption is so significant that the corona and disk are opaque to photons with energies higher than 10 MeV and 1 GeV, respectively.


\section{conclusion and discussion} 
In this letter, we construct an outflow-cloud interaction model to explain the observed 1-30 TeV neutrino data for NGC 1068 and explore the potential suppression of the $\gamma$-ray flux. Protons in outflows that originated from SMBH are going to be accelerated to a maximum energy of 100 TeV by the BS formed outside the cloud. Then those protons will have $p\gamma$ interactions with background photon fields of corona and disk, and have $pp$ interactions with the outflow gas, generating a series of products, including $\gamma$-rays and neutrinos. We find two cases that may explain the observed neutrino data. In Case 1, even though both the $p\gamma$ and $pp$ interactions have contributions to the total neutrino flux, we find that the $p\gamma $ process has the main contribution to the TeV neutrino flux. In Case 2, the $p\gamma$ interaction is no longer important to the neutrino flux, while the $pp$ interaction dominates and can explain the data nicely. As for $\gamma$-rays, due to the high-density photon filed in the corona and disk, the produced $\gamma$-rays will soon be absorbed by the $\gamma\gamma$ attenuation. Therefore, the observed TeV $\gamma$-ray flux is much lower than the neutrino one.


In the outflow-cloud interaction model, whether $p\gamma$ or $pp$ dominates depends on the outflow density. For a specific corona photon field, the outflow density directly determines the $pp$ interaction efficiency exceeding the $p\gamma$ efficiency or not. Therefore, for the $pp$ interaction dominant case, a larger kinetic luminosity of outflow is generally required, which may challenge the total energy budget of the SMBH-disk system. It can be checked based on the future better constraint on the total energy budget of NGC 1068. In addition, some parameters, such as cloud density $n_{c}$, magnetic field strength, and proton spectral index, are still unknown. Thus, further information and limitations are still needed to constrain our parameter space.


\cite{inoue:hal-03373344} also considered the surrounding clouds to contribute the neutrino production through the $pp$ interactions. The difference is that high-energy protons are accelerated by the individual outflow and then enter the cloud to interact with the cloud gas. In our scenario, the outflow will inevitably collide with the clouds, forming the BS and CS. Protons can be effectively accelerated by the BS and advected away with the downstream shocked materials.

Recently, other nearby Seyfert galaxies have been found with characteristics similar to NGC 1068, e.g., NGC 4151, showing a much lower $\gamma$-ray flux than the neutrino flux~\citep{2024ApJ...961L..34M,2024arXiv240606684A}. This may imply that high-energy neutrinos tend to be produced in the environment where $\gamma$ -rays can be suppressed~\citep{PhysRevLett.125.011101}. The central region of AGN is definitely an ideal place. Future deeper observations of these sources with next-generation neutrino telescopes, such as IceCube-Gen2 and Huge Underwater high-energy Neutrino Telescope (HUNT)~\citep{Huang:2023R8}, are crucial for discriminating the diverse models.

\begin{acknowledgments}
We acknowledge support from the National Natural Science Foundation of China under grant No.12003007 and the Fundamental Research Funds for the Central Universities (No. 2020kfyXJJS039).
\end{acknowledgments}

\bibliographystyle{aasjournal}
\bibliography{reference}{}

\end{document}